\documentclass[twocolumn, 10pt]{article}

\usepackage[utf8]{inputenc}
\usepackage[pdftex]{graphicx}
\graphicspath{{figures/}{figures/meth/}{figures/results/}}
\DeclareGraphicsExtensions{.eps,.pdf,.jpeg,.png}

\usepackage[cmex10]{amsmath}
\usepackage{amsfonts}
\usepackage{amsthm}

\usepackage{algorithmic}
\usepackage[caption=false,font=footnotesize]{subfig}
\usepackage{ctable}
\usepackage{authblk}
\usepackage{url}

\newtheorem{definition}{Definition}
\newtheorem{proposition}{Proposition}

\renewenvironment{abstract}{%
\hfill\begin{minipage}{0.99\textwidth}
\rule{\textwidth}{1pt}}
{\par\noindent\rule{\textwidth}{1pt}\end{minipage}\vspace{10mm}}

\makeatletter
\renewcommand\@maketitle{%
\hfill
\begin{minipage}{0.95\textwidth}
\vskip 2em
\let\footnote\thanks 
{\LARGE \textbf \@title \par }
\vskip 1.5em
{\large \@author \par}
\end{minipage}
\vskip 1em \par
}
\makeatother

\begin{document}

\title{An Analytical Model for Loc/ID Mappings Caches}

\author[1]{Florin Coras}
\author[1]{Jordi Domingo-Pascual}
\author[2]{Darrel Lewis}
\author[1]{Albert Cabellos-Aparicio}
\affil[1]{\small Universitat Polit\'ecnica de Catalunya (BarcelonaTECH), Barcelona, Spain}
\affil[2]{\small Cisco Systems, San Jose, CA, USA}

\date{}

\twocolumn[
    
\maketitle 
    \begin{@twocolumnfalse}
        \begin{abstract} 
            \bf
            Concerns regarding the scalability of the inter-domain routing have
            encouraged researchers to start elaborating a more robust Internet
            architecture. While consensus on the exact form of the solution is yet to
            be found, the need for a semantic decoupling of a node's \emph{location}
            and \emph{identity} is generally accepted as a promising way forward.
            However, this typically requires the use of caches
            that store temporal bindings between the two namespaces, to avoid
            hampering router packet forwarding speeds. 
            In this article, we propose a methodology for an analytical analysis of
            cache performance that relies on the working-set theory. We first identify
            the conditions that network traffic must comply with for the theory to be
            applicable and then develop a model that predicts average cache miss rates
            relying on easily measurable traffic parameters. We validate the result
            by emulation, using real packet traces collected at the egress points of a
            campus and an academic network. 
            To prove its versatility, we extend the model to consider cache polluting user traffic and
            observe that simple, low intensity attacks drastically reduce performance,
            whereby manufacturers should either overprovision router memory or implement
            more complex cache eviction policies.

        \end{abstract}
\end{@twocolumnfalse}
]

\section{Introduction}\label{sec:intro}

The growing consensus within the network community today is that a semantic
separation of the \emph{identity} and \emph{location} information, currently
overloaded in IP addresses, would solve the limitations of the routing
infrastructure~\cite{huston:potaroo,rfc4984}. Such architectural update
involves network infrastructure upgrades in few, selected, network points like
border routers, where the linking of the two namespaces must be performed.
But, more importantly, it also entails the more challenging implementation of a
\emph{mapping-system} able to distribute dynamic location-identity bindings
(\emph{mappings}) to globally distributed routers.    

In this context, routers generally retrieve mappings on user demand, as
opposed to proactively fetching them. This is done such that the amount of
memory a router requires to participate in the system does not grow with
identifier space, as is the case today, but is instead dependent on the packet
level traffic the router processes. As a result, to diminish retrieval times,
increase packet forwarding speed and to protect the mapping system from floods
of resolution requests, routers are provisioned with mappings caches
(\emph{map-caches}) that temporarily store in use bindings. 


Although caches placed between processor and main memory, in operating systems
or in web proxies are well studied~\cite{agarwal:cmodel, breslau:webcache,
rizzo:proxycache}, route and mappings caches have yet to be thoroughly
analyzed. A considerable number of experiments have empirically evaluated
map-cache performance, however they are mainly focused on providing a
circumstantial description of cache behavior, that is, for particular cache
configurations and network traffic traces, as opposed to a general
one~\cite{iannone:lcache, jkim:lcache, kim:rcaching, jakab:lisp-tree,
zhang:lcache}. Typically, these results yield accurate estimates for cache
performance but unfortunately cannot be extrapolated to provide rough
projections or bounds on cache behavior for workloads with different characteristics; nor can
they provide insight into what traffic properties influence cache performance
and to what degree.
Answering such questions would not only be a first important step towards
understanding the overall performance of the mapping-system, but would also 
provide a quick way of gauging the expected map-cache performance of any
network domain. 




In this paper, we present an analytical model that, to the best of our
knowledge, constitutes the first theoretical framework for map-cache
performance analysis and provisioning. The model relies on coarse traffic
parameters and aims to be applicable to a wide range of scenarios. In
particular, we first show how the working-set theory~\cite{denning:ws_model} may be
used to estimate simple parameters that characterize the intrinsic locality of network
traffic and thereafter explain how they can be leveraged to link cache size and miss rate. 
The underlying assumption that enables the analysis is that traffic can be
approximated as having a stationary generating process. We find
stationarity to hold for real network traffic, and, to facilitate the use of the model,
we propose a simple methodology that tests for it in network traces.
Finally, we validate the result by emulation, using packet traces collected at
the edges of a campus and an academic network. Part of these results have been
previously presented in~\cite{coras:lcache_n}.



Another contribution of this paper is that we exploit the model to $(i)$
perform an in-depth, over time analysis of cache performance for our datasets
and $(ii)$ study the security of the map-cache by evaluating its vulnerability
to scanning attacks. Our findings show that miss rate decreases at an
accelerated pace with cache size and eventually settles to a power-law
decrease. As a result, even for a relative small size, $10\%$ of the
Internet's aggregated routing table, the performance is still acceptable,
below $0.2\%$ packet miss-rate.
On the other hand, we also find that even simple and low
intensity attacks may compromise performance and that increasing
the cache size does not alleviate the problem while the map-cache cannot
accommodate the whole destination address space. Thereby, to achieve acceptable
forwarding speeds, manufacturers should either overprovision router memory or
design efficient cache management algorithms. 


For the sake of clarity, we focus our analysis on the performance of LISP
map-cache (see Section~\ref{sec:background}). Nevertheless, the results are
relevant for other architectures inspired by the location/identity split
paradigm, including those like ILNP~\cite{RFC6740} that use DNS as their
mapping system, since the equations could be used to approximate DNS resolver
caching performance. Moreover, the cache models could be applied to
route-caching and scalability techniques that focus on shrinking routing
tables to extend router lifetimes~\cite{ballani:viaggre}.

The rest of the paper is structured as follows. Section~\ref{sec:background}
provides an overview of the location/identity split paradigm and one of its
most successful implementations, LISP. In Section~\ref{sec:cmodel} we first
introduce the cache model problem and then present our analytical model and
its extension that accommodates for cache pollution attacks.
Section~\ref{sec:evaluation} describes the methodology used and the results
that validate the two models.  Section~\ref{sec:discussion} discusses the
predictions of our results when applied to our datasets and some design
guidelines to improve cache performance. Section~\ref{sec:rw} presents the
related work. Finally, Section~\ref{sec:conclusions} concludes the paper.

\section{Location/Identity Split Background}\label{sec:background}
Prompted by the growing BGP routing tables and associated churn, researchers
in both academia and operational community have spent recent years (arguably
the last couple of decades) trying to improve the scalability of the
Internet's architecture. The factors supporting the routing table's growth are
partly organic in nature, as new domains are continuously added to the
topology however, they are also closely related to current operational
practices~\cite{rfc4984}. In this sense, multihoming, traffic engineering and
allocations of provider-independent prefixes are the main drivers behind the
prefix de-aggregation that drives the increase of the routing tables. On the
one hand, this effect is detrimental because it leads to engineering
limitations, as the growth rate of the routing table surpasses that of its
supporting technology, but it is also harmful to network operators' business
since the added capital expenses result in less cost-effective
networks~\cite{rfc4984}. 

It has been long hinted at and today it is generally accepted that the origin
of this architectural inflexibility can be tracked down to the semantic
overloading of the IP addresses with both \emph{location} and \emph{identity}
information~\cite{rfc1498}. Therefore, their separation, typically referred to
as a \emph{loc/id split}, has been proposed by many solution that aim to
mitigate the routing problems~\cite{rfc6115} but also by those aiming to
integrate new features~\cite{misseri:diversity}. Apart from infrastructure upgrades,
these architectures also require the deployment of a mapping-system for
linking the two new namespaces and often rely on caches to limit
router memory requirements and improve forwarding speeds. 

Given that the loc/id split paradigm only outlines a set of principles, for
completeness, in what follows we use the Locator/ID Separation Protocol
(LISP)~\cite{rfc6830, saucez:lisp}, to illustrate the operation of a fully-fledged
loc/id architecture. LISP is one of the most successful loc/id split
implementations to date and counts with the support of a sizable
community~\cite{lisp:testbed} dedicated to its development. 

\begin{figure}[t]
    \centering
    \includegraphics[width=80mm,keepaspectratio=true]{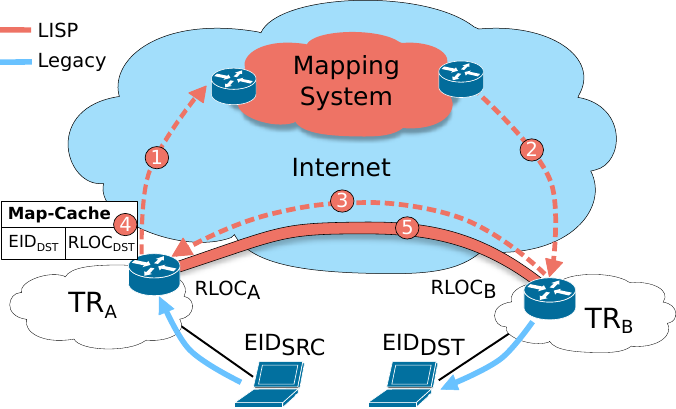}
    \caption{Example packet exchange between $EID_{SRC}$ and $EID_{DST}$ with
    LISP. Following intra-domain routing, packets reach $TR_{A}$ which, not
    having a prior map-cache entry, obtains a mapping binding $EID_{DST}$ to
    $RLOC_{B}$ from the mapping-system (steps 1-3). $TR_{A}$ stores the
    mapping in the map-cache (step 4) and then encapsulates and forwards the
    packet to $RLOC_{B}$  over the Internet's core (step 5). $TR_{B}$
    decapsulates the packets and forwards them to their intended destination. }
    \label{fig:lisp-arch}
\end{figure}

Prior to forwarding a host generated packet (see Fig.~\ref{fig:lisp-arch}), a
LISP router maps the destination address, an end-host identifier (EID), to a
corresponding destination routing locator (RLOC). This is achieved by first
looking up the destination in the local mappings cache, the map-cache, and, if
this fails, by requesting the mapping to a mapping
system~\cite{draft:lisp-ddt,jakab:lisp-tree, menth:firms}. So, instead of
proactively storing bindings for all the possible identifiers in the Internet,
LISP routers request and store them in accordance to the packet level traffic
they process. This ensures that map-cache size is independent of the identity
namespace size and only dependent on local user needs. Once a mapping is
obtained, the border router encapsulates and tunnels~\cite{rfc1955} the packet
from source edge to corresponding destination edge tunnel router (TR), where
the packet is decapsulated and forwarded to its intended destination.

Stale map-cache entries are avoided with the help of timeouts, called \emph{time to
live} (TTL), that mappings carry as attributes, while consistency is ensured
by proactive LISP mechanisms which allow the owner of an updated mapping to
inform its peers of the change. 

The map-cache is most efficient in situations when destination EIDs present
high temporal locality and its size depends on the visited destinations set
size. As a result, performance depends entirely on map-cache provisioned size,
traffic characteristics and the eviction policy set in place.  We dedicate the
remainder of the paper to finding and evaluating the relation between
parameters like cache size, performance and locality of traffic.

\section{Cache Performance}\label{sec:cmodel}
This section presents the theoretical background and analytical methodology
used to model map-cache performance. After briefly introducing the working-set
theory we formalize the cache modeling problem and show that the working-set
is suitable for the analysis of our network traces. The results enable us to
derive a cache model and by extension one that accounts for cache polluting
attacks.


\subsection{Problem Definition}\label{ssec:ws_theory}
For operating systems, a general resource allocation treatment was possible
after it was observed that programs often obey the so called \emph{principle
of locality}. The property arises from the empirical observation that programs
favor only a subset of their information at a given time, so they may be
efficiently run only with a fraction of their total data and instruction code.
It was shown that if the subset, called the program's \emph{locality}, can be
computed, the policy of keeping the locality in memory is an optimal or near
optimal memory management policy~\cite{spirn:locality}.

Based on previous results~\cite{iannone:lcache, kim:rcaching, jkim:lcache}, we
argue that prefix-level network traffic roughly obeys the same principle
because of, among others, skewed destination popularity distributions and flow
burstiness in time, as they gives rise to temporal locality, but also due to
aggregation, i.e., multiplexing of a large number of user flows, which leads
to a form of geographical locality. We therefore evaluate the feasibility of
in-router \emph{prefix/mappings caching} by analyzing the locality of network
traffic. Since we want to avoid any assumptions regarding the structure of the
process generating the network traffic, we opt in our evaluation for the
working-set model of locality. For a list of other locality models
see~\cite{spirn:locality}.  

\begin{figure}[t!]
    \centering
    \includegraphics[width=80mm,keepaspectratio=true]{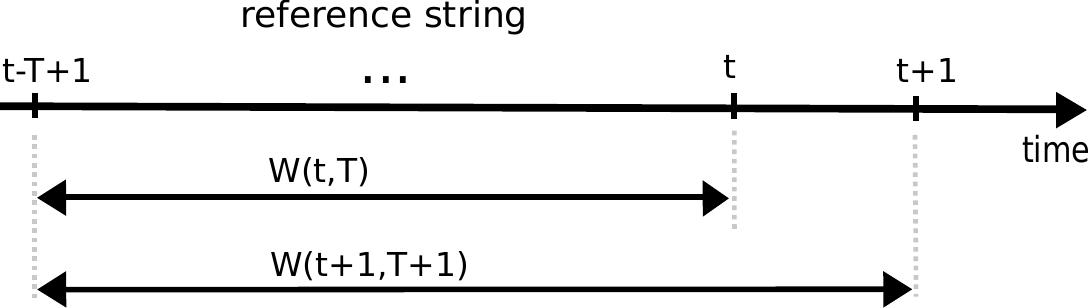}
    \caption{The Working-Set Model. At time $t$, $W(t,T)$ is the set of units
    recently referenced in the time window $T$, i.e., during the interval $[t-T+1,t]$. }
    \label{fig:ws_model}
\end{figure} 

Next we provide a summary of the working-set terminology.  For brevity, we use
the term \emph{unit of reference}, or simply \emph{unit}, as a substitute for
the referenced object (e.g., prefixes); and \emph{reference set} to
represent the set of all referenced units. Then, considering a \emph{reference
set} N, we define a \emph{reference string} as the sequence $ \rho=
r_{1}r_{2}\dots r_{i}\dots$ where each unit $r_{i}\in N$. If $t$ is a measure
of time in \emph{units}, then we can state:

\begin{definition}
Given a reference string, the working-set $W(t,T)$ is the set of distinct
\emph{units} that have been referenced among the $T$ most recent references,
or in the interval $[t-T+1, t]$.
\end{definition}

A graphical depiction can be found in Figure~\ref{fig:ws_model}. In accordance
with \cite{denning:ws_properties} we refer to \emph{T} as the \emph{window
size} and denote the number of distinct pages in $W(t,T)$, the
\emph{working-set size}, as $w(t,T)$. The \emph{average working-set size},
\emph{s(T)}, measures the growth of the working-set with respect to the size
of the window \emph{T}, extending in the past, but independent of absolute
time \emph{t}. It is defined as:

\begin{equation}
    \label{eq:ws}
    s\left(T\right) = \lim_{k\to\infty}\dfrac{1}{k}\displaystyle\sum_{t=1}^k w(t,T)
\end{equation}

It can be proved that the the \emph{miss rate, m(T)}, which
measures the number of units of reference per unit time returning to the
working-set, is the derivative of the previous function and that the sign
inverted slope of the miss rate function, the second slope of $s(T)$,
represents the \emph{average interreference distance} density function,
$f(T)$. For a broader scope discussion of these properties and complete proofs,
the interested reader is referred to \cite{denning:ws_properties}.

It is important to note that $s(T)$ and $m(T)$, if computable, provide
estimates on the minimum average size of a cache able to hold the working-set,
i.e., the prefixes in the active locality, and its corresponding miss rate
with respect to the number of references processed. Our goal in this paper is to
determine if $m(s)$ exists for real network traffic and if it can be modeled as
simple function without a considerable loss of precision.

\subsection{Network Traffic Locality}\label{ssec:tprop}
\begin{figure*}[t!]
    \centering
    \subfloat[$upc~2009$]{\label{fig:ws1}\includegraphics[width=0.25\textwidth,keepaspectratio=true]{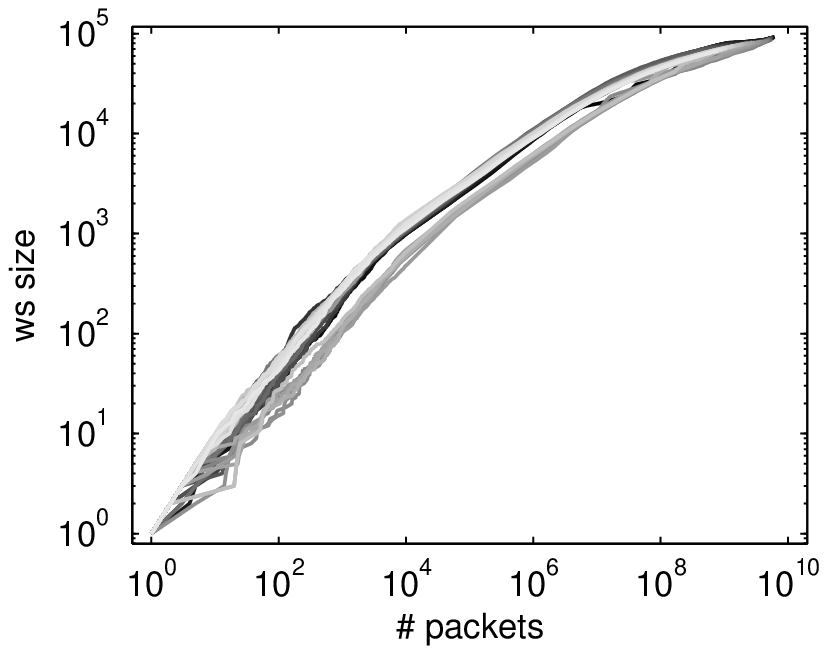}}
    \subfloat[$upc~2011$]{\label{fig:ws2}\includegraphics[width=0.25\textwidth,keepaspectratio=true]{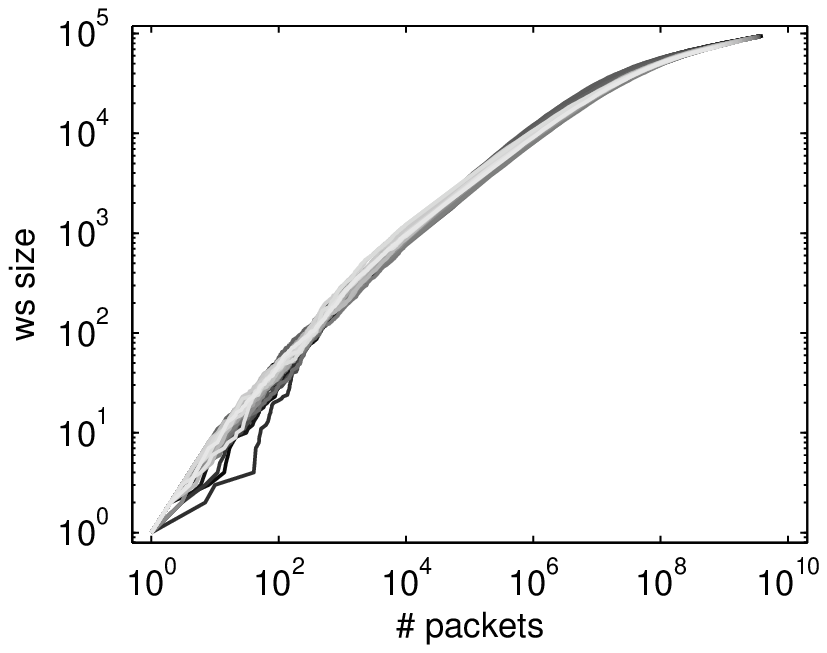}}
    \subfloat[$upc~2012$]{\label{fig:ws4}\includegraphics[width=0.25\textwidth,keepaspectratio=true]{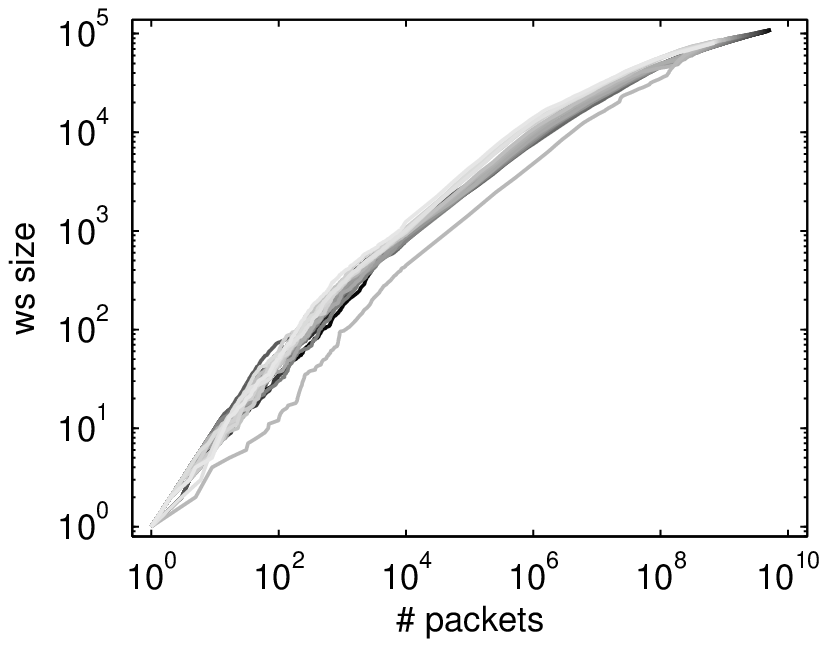}}
    \subfloat[$cesca~2013$]{\label{fig:ws3}\includegraphics[width=0.25\textwidth,keepaspectratio=true]{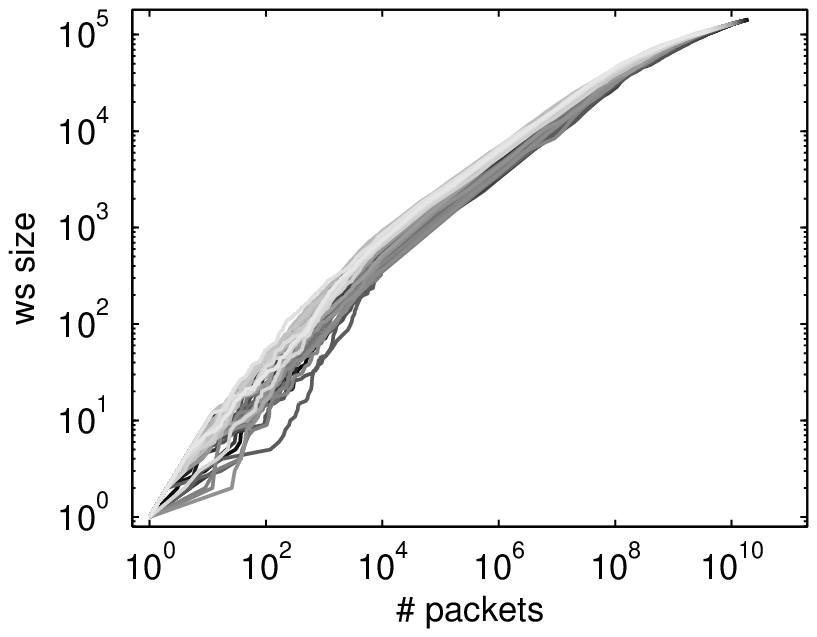}}
    \caption{Empirical working-set curves with starting times spaced by $30$
    min and evenly distributed over a day. Closeness of color nuance reflects closeness of start time. Notice their tendency to cluster. }
    \label{fig:ws_empirical}
\end{figure*}



As explained by Denning in \cite{denning:ws_properties}, a working-set
analysis of reference strings may be performed only if based on three
constraints that provide for a more rigorous definition for \emph{locality of
reference}:

\begin{enumerate}
 \item  Reference strings are unending
 \item  The stochastic mechanism underlying the generation of a reference string
        is stationary, i.e. independent of absolute time origin.
 \item  For $t>0$, $r_{t}$ and $r_{t+x}$ become uncorrelated as
        $x\rightarrow\infty$
\end{enumerate}

The first, though obviously not fulfillable, introduces an insignificant error
because the reference strings generated by practical programs or traces are
long from a statistical standpoint. The third requires that references become
uncorrelated as the distance goes to infinity. This can usually be asserted as
being true in practice. The most restrictive of the three is the second, which
limits the analysis to a locality where all three constraints, including
stationarity are satisfied. 

In practice however, network traffic reference strings may consists of
sequences of localities when either each is characterized by a distinct stationary
processes or, alternatively, when some present nonstationarities. In such scenarios
the results, like the average working-set size, are only valid within
a locality, and not to be extrapolated for the whole trace. To test for
this possibility and to identify the reference string segments having
different generating processes in network traffic traces, we devise the simple
experimental methodology that follows.

To enable our analysis, since computing the $w(t,T)$ for all acceptable
integer combination of $t$ and $T$ is intractable, we define the
\emph{working-set curve} to be $w(t,T)$ as a function of $T$, when the past
time reference of the working-set is held fixed, i.e., $t-T=cst$. For instance,
considering Figure~\ref{fig:ws_model}, $w(t,T)$ and $w(t+1, T+1)$ are
consecutive points on a working-set curve with start time $t-T$. Then, for a
given trace, we compute multiple empirical destination prefix working-set
curves with start times spanning one day and spaced by a fixed interval.
Intuitively, one would expect that the clustering patterns of the curves
should distinguish between the multitude of generating process. That is,
curves with close start times should have a similar growth shape (cluster),
because they follow a similar sequence of localities, whereas those separated
by larger time lags should behave differently.
More formally stated: 

\begin{proposition} 
    \label{prop:clustering}
    The clustering of the working-set curves, under which $\forall T,
    w(t,T)$ is normally distributed, is equivalent to the stationarity of the
    process generating the reference string.  
\end{proposition}

We provide a proof for the the proposition in the Appendix. In addition, we
empirically confirm the result by using it to determine the stationary processes
embedded in four real network traffic traces. Details regarding the traffic captures
can be found in Section~\ref{ssec:traces}.

For each network trace we computed working-set curves spaced by half an hour.
Figure~\ref{fig:ws_empirical} presents the working-set curves of the four
datasets in log-log plots. It can be noticed that all traces exhibit a strong
clustering and sublinear growth, due to temporal locality. Although, the
number of samples does not allow for an accurate enough testing, we could also
confirm that $\forall T, w(t,T)$ is close to normally distributed by manual
inspection and graphically. Therefore, in light of the previous proposition,
we find that each trace can be considered as generated by a single stationary
process.  


To validate the result, we also independently tested the stationarity of the
process generating the reference string by applying the augmented
Dickey-Fuller unit root test~\cite{banerjee:adf} to the interreference
distance time series. Due to the large size of the dataset we first aggregated
the time series by computing for each window of $10k$ data points the mean and
subsequently applied the test to the resulting mean series. For all traces,
the null hypothesis that the series has a unit root, with the alternative that
the series is stationary, was rejected ($p<0.01$).

Undoubtedly, the most surprising result of the analysis is the stationarity of
the processes generating the four traces. It follows that the average
working-set and the other metrics derived from it characterize a trace in its
entirety. This might seem to be somewhat counter-intuitive if one considers
the nonstationarities of network traffic when analyzed in the time domain.
However, we believe stationarity to be the result of flow multiplexing,
whereby the effect of short term correlations is canceled out by the influence
of destination popularity, as shown in the case of Web traffic
in~\cite{jin:web_tloc}.

Similarly, Kim et al. observed in \cite{kim:rcaching} that the working-set
size for prefixes tends to be highly stable with time for traffic pertaining
to a large ISP. We would like to stress that we do not assume all traffic in
the Internet is generated by stationary processes, i.e., possesses an
approximate time translation invariance of the working-set curves like the one
observed in Figure~\ref{fig:ws_empirical}. And in fact, we require that the
model we develop further be applied only to traces that have this property.





\subsection{Analytical Cache Model}\label{ssec:model}
Because the exact form of the average working-set size, as obtained using
(\ref{eq:ws}), is rather cumbersome to work with, a shorter but approximate
representation would be desirable. With hindsight, one can recognize that the
empirical working-set curves from Figure~\ref{fig:ws_empirical} are piecewise
linear when depicted in log-log scale. This observation enables us to
approximate the average working-set size, $s(u)$, for each trace with respect
to the number of packets $u$, by means of a piecewise linear fit of the
log-log scale plot. We therefore obtain estimates of both the slope, $\alpha$,
and the y-intercept, $\beta$, for all segments. In our results, we limited the
number of segments to just four, however if better fits are desirable, more
segments may be used.  Through conversion to linear scale the average
working-set equation becomes piecewise power law of the type:  

\begin{equation}
    \label{eq:ws_fit}
    s(u)=e^{\beta(u)}u^{\alpha(u)}
\end{equation}

\noindent where, $u$ represents the number of referenced destination prefixes,
or the window size, $s(u)$ is the fitted working-set size function and
$\alpha(u)$, $0<\alpha(u)\le1$, and $\beta(u)\ge0$ are piecewise constant,
decreasing and respectively increasing functions obtained through
fitting. Defined as such, the pair $(\alpha(u),
\beta(u))$ provides a compressed characterization of the temporal
locality present within a trace with respect to time, i.e., number of packets.  

We can estimate the miss rate for a trace by
computing the derivative of $s(u)$ like:

\begin{equation}
    m(u)=e^{\beta(u)}\,\alpha(u)\,u^{\alpha(u)-1}
    \label{eq:mr_est}
\end{equation}

Taking the inverse of (\ref{eq:ws_fit}) and inserting it in
(\ref{eq:mr_est}) we obtain an analytical relation that links the cache size
and the estimated miss rate:

\begin{equation}
    m(s)=e^{\,\beta^{*}(s)/\alpha^{*}(s)}\,\alpha^{*}(s)\,\,s^{1-1/\alpha^{*}(s)}
    \label{eq:mr_vs_cs}
\end{equation}

\noindent
where, $s$ represents the cache size in number of entries and $\alpha^*(s)$
and $\beta^*(s)$ are piecewise constant functions with knees dependent on $s$.
This equation accurately predicts cache performance over longer spans of time,
if $s(u)$ is relatively stable in the considered time frame (more on this in
Section~\ref{sec:discussion}). 

Regarding the type of cache modeled, it is useful to note that the
working-set, $W(t,u)$, generally models a cache that always contains the
$w(t,u)$ most recently referenced units. Then, given that $\forall u, w(t,u)
\sim N(\sigma,\mu^2)$, $W(t,u)$ actually models a cache of size normally
distributed and dependent on $u$. In particular, when $\sigma$ is small, or
goes to zero, it behaves like the size were fixed. Finally, because the
implicit eviction policy requires that entries not referenced in a window of
length $u$ are discarded, for low $\sigma$, like in our traces, the
working-set simulates a cache of fixed size with a LRU eviction policy. 

To summarize, because we find that $\forall u$, the working-set size is
normally distributed and that traffic may be seen as generated by a stationary
process, (\ref{eq:mr_vs_cs}) actually models a cache with LRU eviction and
size dependent on $u$.  We validate the model and the efficiency of LRU in
Section~\ref{ssec:results}.

\subsection{Cache Model for Cache Pollution Attacks}\label{ssec:attack_model}
The model presented in the previous section can be extended to account for
situations when intra-domain users perform EID space scanning attacks
that significantly alter the working-set curves and therefore the shared
map-cache's performance. 
In this work we focus on assessing the damage users can inflict through
data-plane attacks and do not consider control-plane attacks like those
described in~\cite{draft:lisp-threats}.

We define a scanning attack as the situation when one or multiple users, acting
jointly, send packets over a large period of time (e.g., hours), to
destinations having a high probability of not being found in the cache. The
goal would be to either generate cache misses, resulting in control plane
overload or, if the cache is not large enough, to generate cache evictions,
which would affect ongoing flows. For instance, having a list of EID prefixes,
an attack would consist in sending packets with destinations enumerating all
prefixes in the set in a random order, at a certain packet rate. Once all
destinations are exhausted the enumeration would start over.

In what follows we formally define the parameters of the attack.  Let,
$\Omega$ be the EID-prefix set used in the attack and $\Psi$ the network's
visited EID-prefix set. We define the relative attack intensity $\rho$ as the
ratio between the attack packet rate and the legitimate traffic packet rate,
additionally, let the attack overlap $\delta$ be the ratio between the number
of prefixes common to $\Omega$ and $\Psi$ and the cardinality of $\Psi$ thus,
$\delta = |\Omega \cap \Psi|/{|\Psi|}$.

If a network trace with average working-set $s(u)$ is augmented by a scanning
attack of relative attack intensity $\rho$ and overlap $\delta$, the resulting
average working-set becomes:

\begin{equation}
    s_{a}(u+\rho u) = 
        \begin{cases}
            s(u)+\rho u - \dfrac{\delta s(u_{k})}{u_{k}}u, & u<u_{k}\\
            s(u)+|\Omega|-\delta s(u)
        \end{cases} 
    \label{eq:wsa_s1}
\end{equation}

\noindent where $u_{k}= |\Omega|/\rho$ and it represents the number of
legitimate packets after which the attack exhausts all $|\Omega|$ destinations
and the scan restarts. The aggregate working-set has three components. The
first is due to legitimate traffic, $s(u)$, and the second, due to the attack
packets, $\rho u$. However, because the two may overlap, a third component
subtracts the number of shared prefixes. For simplicity, we approximate
the probability of having destinations repeat to be uniform. Thereby, the growth
of the overlap is linear with $u$ up to $u_{k}$, where it reaches a
maximum of $\delta s(u_{k})$ and afterwards linear with $s(u)$. 

After a change of variable and denoting $\tau=1/(1+\rho)$, or the ratio of
legitimate traffic in the trace, $u_k =\dfrac{|\Omega|}{1-\tau}$ and the equation becomes:

\begin{equation}
    s_{a}(u) = 
        \begin{cases}
            s(\tau u)+ \left(1-\tau-\dfrac{\tau\delta s(u_{k})}{u_k} \right)u, u< u_k \\ 
            (1-\delta)s(\tau u) + |\Omega|
        \end{cases} 
    \label{eq:wsa}
\end{equation}

\noindent
Then, the miss as a function of the number of processed packets is:

\begin{eqnarray}
    m_{a}(u) = 
        \begin{cases}
            \tau\,m(\tau u)+\left(1-\tau-\dfrac{\tau\delta s(u_{k})}{u_k} \right), u<u_k\\
            \tau(1-\delta)\,m(\tau u)
        \end{cases} 
    \label{eq:wsad}
\end{eqnarray}

However, in this case the miss rate cannot be represented analytically as a
function of the cache size since $s_a^{\,-1}(u)$ is not expressible in terms of
standard mathematical functions. It can though be computed numerically as a
function $u$, when $s(u)$ is known. Then, given that both $s_a(u)$ and
$m_a(u)$ are known, they suffice to understand the cache's miss
rate as a function of the cache size. 
The resulting model predicts overall cache misses, not only those due to
legitimate traffic. Therefore, it provides an estimate of the control plane
overload, not a data plane performance estimate for legitimate traffic.
We empirically validate the result in Section~\ref{ssec:results}.

\section{Model Validation}\label{sec:evaluation}
In this section, we present the evaluation methodology and the results that
validate our models. We start by describing our datasets and then pass on to
present the ITR emulator employed in the empirical evaluation of the cache.
Last, we compare the empirical results to those predicted by the analytical
models.

\subsection{Datasets}\label{ssec:traces}
Four one-day packet traces were used for the purpose of our experiments.
Three of them have been captured at the 2Gbps link connecting several of our
University's campus networks to the Catalan Research Network (CESCA) and
consist only of egress traffic. UPC campus has more than 36k users. The first
trace dates back to May 26th, 2009 while the other two are from October 19th,
2011 and November 21st, 2012. The fourth trace was captured on January 24th,
2013 at CESCA's 10Gbps link connecting it to the Spanish academic network
(RedIris) and consists of egressing academic traffic. In 2011, CESCA was
providing transit services for $89$ institutions, including the greater part
of public Catalan schools, hospitals and universities. 

Table~\ref{tab:traces} summarizes some of the important properties of our
datasets. First of all, it can be seen that $cesca~2013$, being an aggregate
trace, is about $3.6$ times larger than the most recent UPC trace in terms of
number of packets and packet rate. However, it only contains $1.3$ times more
prefixes. This shows that although the number of users and packets exchanged
is considerably higher, the diversity of the destinations is only slightly
incremented. Out of the UPC traces, $upc~2009$ exhibits a surprisingly high
number of packets but this is explained by the very large packet rates seen
during the active hours of the day. In fact, the average packet rate during
the peak hours was $4.7$ times higher that for the rest of the day. Again,
this difference did not reflect in the number of unique prefixes observed in a
one second window as, on average, we observed just $1.3$ times more prefixes
in the peak hours than during the remainder of the day. These two observations
suggest that higher packet rates, either resulting from larger user sets or
from higher throughput flows, do not increase destination diversity (as
illustrated in Figure~\ref{fig:ws_empirical}) but instead reinforce temporal
locality. In addition, these properties also explain why the working-set
curves for $upc~2009$ present a time-of-day behavior (see
Figure~\ref{fig:ws1}).


\begin{table*}[t]
    \centering
    \caption{Datasets Statistics}
    \label{tab:traces}
    \begin{tabular}{lcccc}
        \toprule
                & \textbf{upc 2009} & \textbf{upc 2011} & \textbf{upc 2012} & \textbf{cesca 2013} \\ \midrule[0.09em]
        Date    & 2009-05-26 & 2011-10-19 & 2012-11-21 & 2013-01-24\\ \midrule 
        Packets & 6.5B & 4.05B & 5.57B & 20B \\ \midrule
        Av. pkt/s & 75312 & 46936 & 64484 & 232063  \\ \midrule
        $\Psi$  & 92801 & 94964 & 109451 & 143775 \\ \midrule
        Av. pref/s & 2323 & 1952 & 2123 & 2560 \\         
        \bottomrule
    \end{tabular}
\end{table*}

\begin{table*}[t]
    \centering
    \caption{Routing Tables Statistics}
    \label{tab:rt}
    \begin{tabular}{lcccc}
        \toprule
                & \textbf{upc 2009} & \textbf{upc 2011} & \textbf{upc 2012} & \textbf{cesca 2013} \\ \midrule[0.09em]
        $\texttt{BGP}_{RT}$ & 288167 & 400655 & 450796 & 455647 \\ \midrule
        $\texttt{BPG}_{\phi}$ & 142154 & 170638 & 213070 & 216272 \\ \midrule
        $\Psi/\texttt{BGP}_{\phi}$ & 0.65 & 0.55 & 0.51 & 0.66 \\
        \bottomrule
    \end{tabular}
\end{table*}

\subsection{Map-Cache Emulator}\label{ssec:simulator}
To evaluate the two models and the effectiveness of the working-set as a tool for
cache performance prediction, we implemented a packet trace based emulator
that mimics basic ITR functionality.

Both for computing the working-sets in Section~\ref{ssec:tprop} and for the
cache performance evaluation, destination IP addresses had to be mapped to their
corresponding prefixes. We considered EID-prefixes to be of BGP-prefix
granularity. For each traffic trace, we linked IP addresses to prefixes using
BGP routing tables ($\texttt{BGP}_{RT}$) downloaded from the RouteViews
archive~\cite{routeviews} that matched the trace's capture date. In
particular, we used collector \emph{route-views4} situated at University of
Oregon. The only preprocessing we performed was to filter out more specific
prefixes. Generally, they are used for traffic engineering purposes but LISP
provides mechanisms for a more efficient management of these operational needs
that do not require EID-prefix de-aggregation. We refer to the resulting list
as $\texttt{BGP}_{\phi}$. Table~\ref{tab:rt} shows the size of the routing
tables used for each trace and provides the proportion of prefixes seen within
each trace out of the total registered in the filtered routing table,
$\Psi/\texttt{BGP}_{\phi}$. It may be seen that, as the ratio is always higher
that $0.5$, more than half of the possible destination prefixes are visited in
one day for all traces.

For each packet processed, the emulator maps the destination IP address to a
prefix in $\texttt{BGP}_{\phi}$. If this prefix is already stored in the ITR's
cache, its cache entry is updated and the emulator continues with the next
packet. Should the prefix not yet be stored in the cache, two possibilities
arise. First, if the cache is not full, the destination prefix is stored in
and the processing proceeds to the next packet. Second, if the cache is full,
an entry is evicted, the new prefix is stored in and then the emulator moves
to the next packet. The entry to be evicted is chosen according to the LRU
eviction policy. We use LRU because, as mentioned in Section~\ref{ssec:model},
its performance should be close to optimal due to the stationarity of the
trace generating process. Accordingly, the performance of the cache should be
appropriately described by (\ref{eq:mr_vs_cs}).

Scanning attacks are emulated by uniform insertion of attack packets
in-between those pertaining to the processed traffic trace, according to the
relative attack intensity $\rho$. The number of attack addresses generated
depends on $\delta$, the overlap between $\Omega$ and $\Psi$, and the number
of destinations in $\texttt{BGP}_{\phi}$. Note that $\Omega \subseteq
\texttt{BGP}_{\phi}$ and $\Psi\subseteq\texttt{BGP}_{\phi}$. Therefore,
supposing the attack maximizes number of addresses used, to increase
effectiveness, $|\Omega|=|\texttt{BGP}_{\phi}-\Psi|+\delta|\Psi|$. In particular,
when no overlap exists, we generate $|\Omega| = |\texttt{BGP}_{\phi} - \Psi|$
new destination addresses while for a full overlap, the attack consists of
$|\Omega|=|\texttt{BGP}_{\phi}|$ addresses. If $\delta\neq0$, the addresses
used in the attack and part of $\Psi$ are uniformly distributed among those
part of $\texttt{BGP}_{\phi} - \Psi$.

\begin{figure*}[t!]
    \centering
    \subfloat[$upc~2009$]{\label{fig:miss1}\includegraphics[width=0.25\textwidth,keepaspectratio=true]{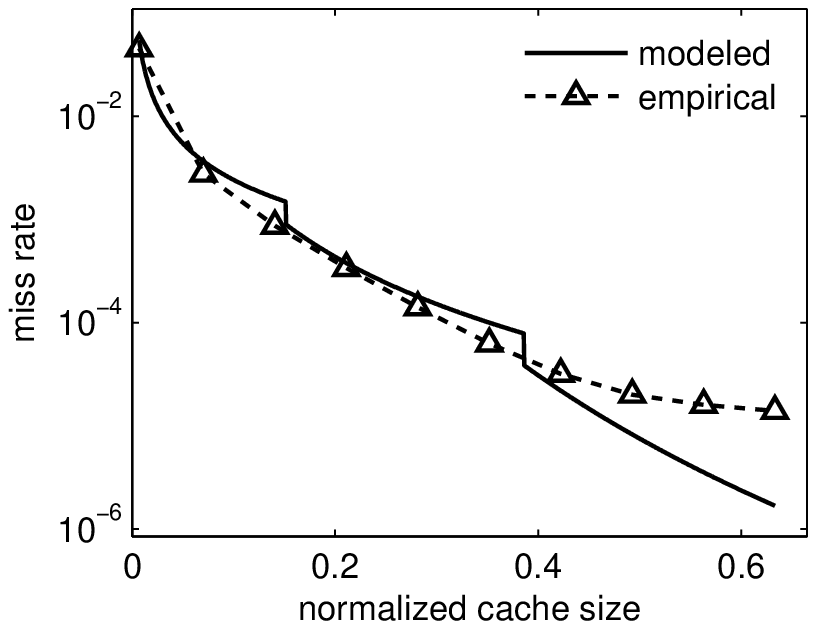}}
    \subfloat[$upc~2011$]{\label{fig:miss2}\includegraphics[width=0.25\textwidth,keepaspectratio=true]{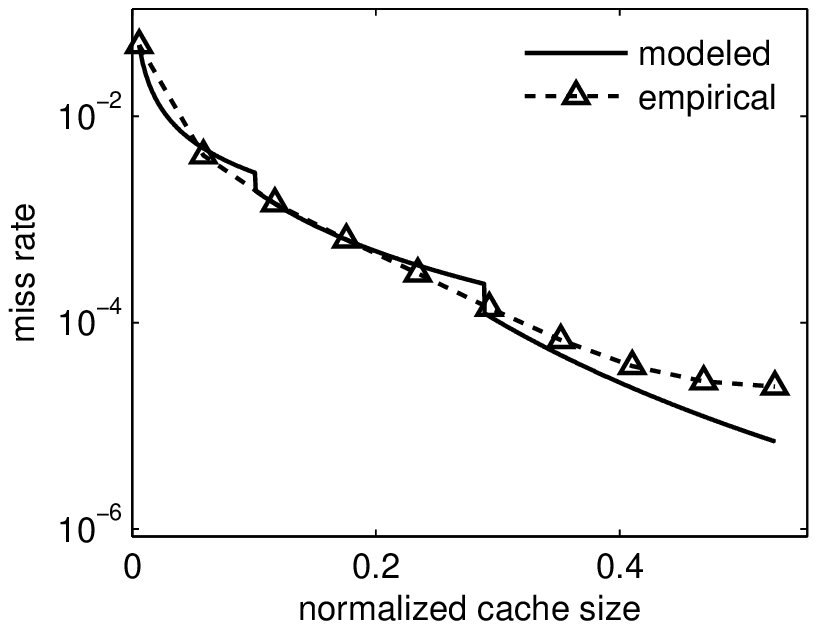}}
    \subfloat[$upc~2012$]{\label{fig:miss3}\includegraphics[width=0.25\textwidth,keepaspectratio=true]{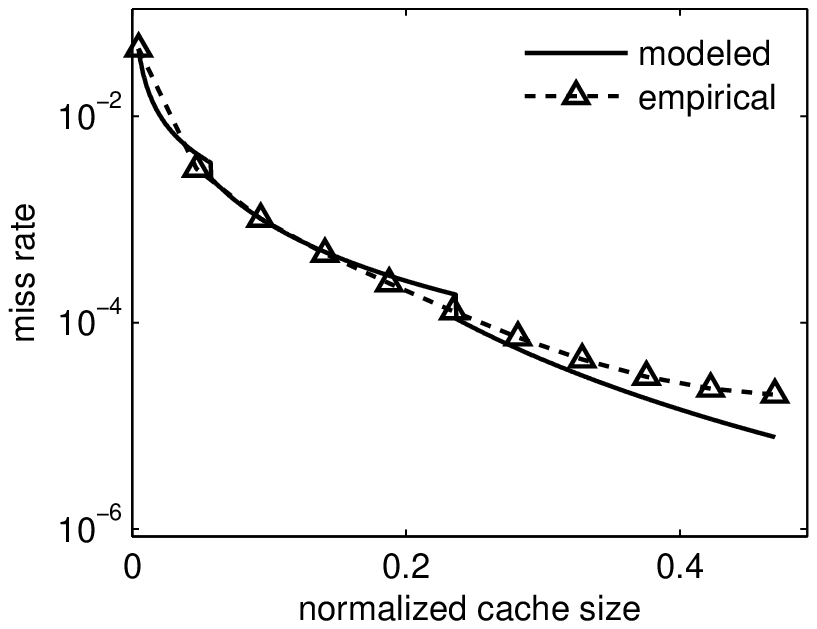}}
    \subfloat[$cesca~2013$]{\label{fig:miss4}\includegraphics[width=0.25\textwidth,keepaspectratio=true]{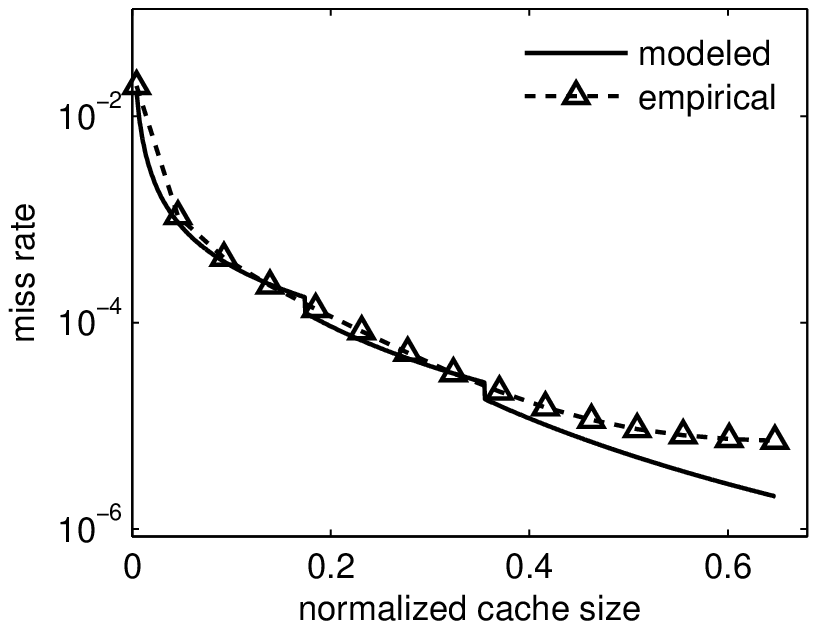}}
    \caption{Comparison between empirical and modeled miss rates, as estimated
        by~(\ref{eq:mr_vs_cs}), for normal cache operation.  }
    \label{fig:miss_fit}
\end{figure*}

\subsection{Comparison of Analytical and Empirical Results}\label{ssec:results}
To validate the models we use the emulator to estimate empirical cache miss
rate for several cache sizes. Figure~\ref{fig:miss_fit} presents a comparison
of the empirical and predicted results for normal traffic, where cache size is
normalized with corresponding $\texttt{BGP}_{\phi}$ routing table size (see
Table~\ref{tab:rt}). It may be seen that, typically, the absolute error
is negligible despite the discontinuities of the model, which are due to
the piecewise fitting.  
Further, the equation appropriately predicts that performance stays acceptable,
even for small cache sizes, fact also observed by
\cite{iannone:lcache,jkim:lcache,jakab:lisp-tree}. The result is even more
remarkable as we never remove stale entries in our emulator, i.e., we consider
TTL to be infinite. 

The figures present the cache miss rate only up to a fraction of
$\texttt{BGP}_{\phi}$, the one associated to $|\Psi|$, because the growth of
$s(u)$ cannot be extrapolated after this point. Since the working-set grows
slower for larger $u$, as $\alpha(u)$ is a strictly decreasing function, potentially
much longer traces would be needed to enable inference about larger cache
sizes. In fact, given that only part of the whole prefix space may be visited
by the clients of a stub network, even for longer traces the analysis may be
limited to a cache size lower than $|\texttt{BGP}_{\phi}|$. 

There are two limitations to the precision of our analysis for large $u$
values. First, as cache size increases and approaches $|\Psi|$, the accuracy
of the prediction diminishes. This is explained by the 24-hour length of the
traces, whereby there are few working-set curves that span close to a whole
day and thus grow to reach the maximum number of destination prefixes. Recall
that the start times for the working-set curves span the whole trace and are
spaced by $30$ minutes, so the last curves consist of few packets. As a
consequence, $s(u)$ is estimated using a reduced number of points, i.e., with
a lower precision, at the higher end. To counter this effect, we compute
average working-sets of slightly diminished length. The second limitation is
the bias of our emulation results for large cache sizes. Caches whose sizes
are close to $|\Psi|$ fill only once the traces are processed almost in their
entirety. Due to this cold-start effect, cache are exposed to a low number of
hits up to the end of the traces. As these hits do not manage to outweigh the
misses generated during the cache fill up, the miss rate of the emulator for
large cache sizes is slightly overestimated. Nevertheless, despite these
limitations, it may be seen that the results still yield a good fit for large
$u$.  

\begin{figure}[h!]
    \centering
    \subfloat[No overlap ($\delta=0$), $\Omega =
        \texttt{BGP}_{\phi}-\Psi$]{\label{fig:attack_reduced}\includegraphics[width=0.45\textwidth,keepaspectratio=true]{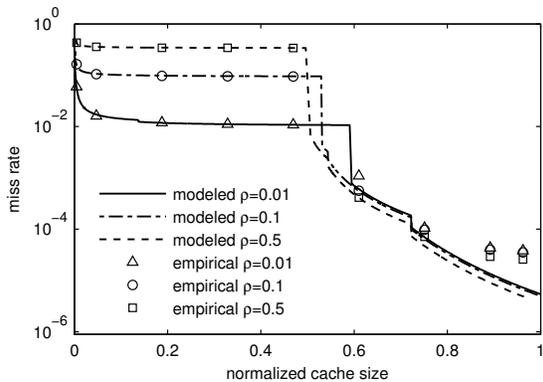}}\\
    \subfloat[Complete overalp ($\delta=1$), $\Omega=\texttt{BGP}_{\phi}$]{\label{fig:attack_full}\includegraphics[width=0.45\textwidth,keepaspectratio=true]{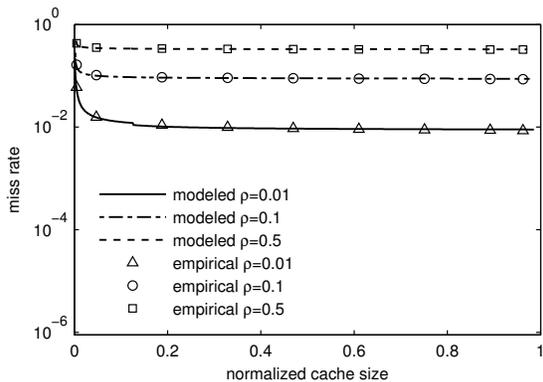}}
    \caption{Comparison between empirical and modeled miss rates, as estimated
        by~(\ref{eq:wsa}) and~(\ref{eq:wsad}), under scanning attacks with attack intensity $\rho \in\{0.01, 0.1, 0.5\}$ for
        \emph{upc 2012}.}
    \label{fig:attack_fit}
\end{figure}

We also validate our cache model that accounts for scanning attacks
considering as scenarios a complete attack overlap or one that is zero.
Thereby, the attackers may use as attack prefix set either the whole
EID-prefix space or just the part not visited by the attacked network's
clients. In the latter case, if the cache is not large enough to hold all
prefixes, all attack packets would generate a cache miss. However, note that
building such a prefix set would require full knowledge about the network's
traffic. In the former case, some packet destinations may generate cache hits
but $|\Omega|$ may potentially be much larger and this could prove beneficial
to the attacker. In light of these properties we consider the two attacks as
\emph{worst case} scenarios, from the attacked network's perspective, for the
situations when attackers respectively have or do not have knowledge about the
attacked network's traffic.

Figure~\ref{fig:attack_fit} compares the analytical and empirical results for
the cache miss rate, when $\delta \in\{0,1\}$ and $\rho \in\{0.01, 0.1,
0.5\}$. We present the results just for \emph{upc 2012} since those for the
other three traces are similar. It may be observed that for both no
overlap and complete overlap the results exhibit little absolute error. In
particular, for $\delta=0$ when cache size is larger and miss rates are less
than $10^{-4}$ the errors are more significant. As in the case of the cache
model, this is explained by the trace length and the reduced number of points
used in estimating the higher end of $s(u)$. The effect is not noticeable for
$\delta=1$ because $s_a(u)$ reaches its maximum (saturates) for low values of
$u$ due to the larger attack set. Consequently, the fit is very good along the
whole spectrum of cache sizes. Once the cache size reaches
$|\texttt{BGP}_{\phi}|$ the miss rate becomes $0$ since there are no more
destination prefixes outside those already present in the cache to generate a
miss.

\section{Discussion}\label{sec:discussion}
In this section we discuss the results and predictions of our models regarding 
map-cache performance for both normal and malicious traffic. We also
discuss possible avenues to diminishing the effect of cache attacks.

The results we obtained are relevant only when reasoned about jointly with the
traffic traces used in the analysis. However, the diversity of our data sets
and previous results from Kim~\cite{kim:rcaching},
Iannone~\cite{iannone:lcache} and Kim~\cite{jkim:lcache} suggest that the
properties uncovered are not the expression of isolated user and network
behavior.

\subsection{Cache Model Results and Predictions} 

Under the condition of a stationary generating process or, equivalently, the
approximate time translation invariance of the working-set curves, our
methodology enables the estimation of the time invariant piecewise functions
$\alpha(u)$ and $\beta(u)$ that characterize the locality of a network traffic
trace from the average working-set size $s(u)$. This further facilitates the
following two findings. First, due to the low variance of $s(u)$ and
experimentally proven good performance, we can now recommend the use of the
LRU eviction policy for LISP caches. Second, in such situation,
(\ref{eq:mr_vs_cs}) may be used to dimension the cache sizes in operational
environments, according to the desired miss rate. The prediction of its
mathematical expression, considering that $\alpha^*(c)\to0$ when $c$
increases, is that miss rate decreases at an accelerated pace with cache size
and finally settles to a power-law decrease. This may also be observed in
Figure~\ref{fig:miss_fit} where at each discontinuity point the function
switches to a faster decreasing curve. Of course, the speed of the decrease
depends on the degree of locality present in the trace. Overall the equation
indicates that cache sizes need not be very large for obtaining good
performance. For instance, having a cache of size $10\%$ of $BGP_{\phi}$,
about $14k-21k$ entries for UPC traces and $21k$ for the CESCA one, would
result in a miss rate of approximately $0.09\%-0.2\%$ and respectively
$0.03\%$.

\begin{figure}[t!]
    \centering
    \includegraphics{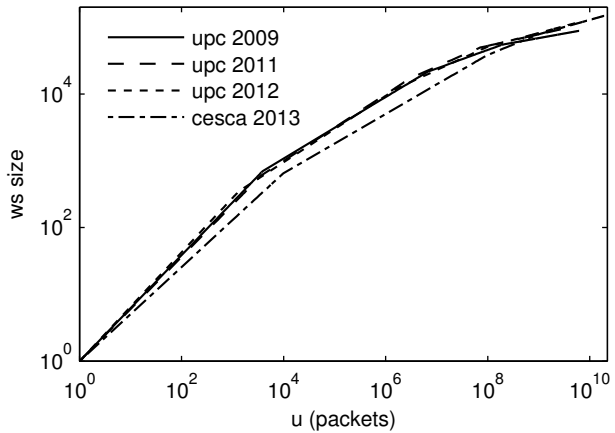}
    \caption{Evolution of $s(u)$ over the years. The shape appears to be influenced
    by the number of users and the destination popularity distribution.}
    \label{fig:ws_evolution}
\end{figure}

In this context, an important point would be to determine the extent of time
over which the results and predictions hold. Figure~\ref{fig:ws_evolution}
provides a coarse answer for the particular case of the traffic used in our
analysis. First, considering the UPC traces, it may be observed that over a
span of tree and a half years, the average working-set is rather stable when
size is less than about $50k$ prefixes. In fact, the $s(u)$ curves of
$upc~2011$ and $upc~2012$ are very similar, independent of $u$ value, while
the one for $upc~2009$ exhibits a lower slope for $u>100M$ packets. This might
appear to be in agreement with the relative, year-over-year, increase of
$BGP_{\phi}$. But the relative differences between the values of $s(u)$ and
the increase of $BGP_{\phi}$ are not directly related since the growth from
$2009$ to $2011$ was smaller than the one from $2011$ to $2012$. This is to be
expected because many of the new prefixes, resulting from the sustained growth
of the Internet's edge~\cite{cittadini:evolution}, may never become
destinations for users of other edge networks. So, a direct relation between
the increment of the routing table size and that of $s(u)$ should not
necessarily be expected. Instead, given the good overlap for lower values of
$u$, prefix popularity distribution should be more relevant to the shape of
the working-set than the absolute number of destination. Second, the
comparison between $cesca~2013$ and the UPC traces reveals that larger user sets,
and implicitly higher traffic rates, result in a slightly slower growing $s(u)$.
In fact, the only noticeable difference is at short time scales, where 
the larger trace has a smaller slope. This could be explained by a change in
the destination popularity distribution, as $cesca 2013$ aggregates
more types of user profiles, but also by a shift in the short-term temporal
correlation. However, considering the large number of users, their
synchronization at short time scales seems rather unlikely.

Then, although apparently stable over relatively long time spans, the shape of
$s(u)$ seems to be influenced by non trivial interactions between the number
of clients the prefix popularity distribution and possibly other unexplored
factors. We are not ready to model these interactions here. Thus, our cautious
inference is that the average working-set should be stable over time, if the
number of clients and the popularity distribution are relatively stable.

Despite not being indicated by our measurements, it may be finally proven that
the variability with time of $s(u)$ is highly dependent on properties of the
network being measured, themselves time dependent. Should this be the case, the
methodology we develop in this paper is still valuable for the analysis of
cache performance if not for long term provisioning of caches. 

\subsection{Cache Poisoning and Management Strategies}

We compared the effect of scanning attacks with zero or complete overlap to
identify the one most damaging and to quantify their effect on cache
performance. Associated results were presented in Figure~\ref{fig:attack_fit}. 

If cache sizes are small, both attacks results in very high cache miss rate,
including for $\rho=0.01$, when the attack has a rate of only $644$ packets
per second. In this range, miss rate is almost independent of attack overlap,
only slightly higher for $\delta=0$ due to the informed selection of attack
address space. However, for $\delta=0$ the cache performance is much improved
after a certain threshold is passed whereas for $\delta=1$ it barely changes
up to when cache size becomes $\texttt{BGP}_{\phi}$ and miss rate drops to
$0$.  In other words, the non overlapping attack may be absorbed with larger
cache sizes while the overlapping one no. Perhaps counter-intuitively,
overlapping attacks are more damaging against a map-cache. They are easier to
generate, as they do not require prior knowledge about the attacked network.
But, they are also harder to defend against since, after a certain point and
for a wide ranges of values, increasing the cache sizes does not yield much
improved performance if less than $\texttt{BGP}_{\phi}$. 

Arguably, the most worrisome result we observe is the rather high miss rate
which barely drops under $0.01$, even for $\rho=0.01$ and only when the normalized
cache size is higher than $0.2$. As a comparison, under normal operation this
miss rate would be obtained with a normalized cache size of about $0.02$, an
order of magnitude less. Therefore, considering the high packet throughput of
border routers, some more complex cache management strategies should be set in
place to avoid hundreds to thousands of packet drops per second. 

One possible first step to circumventing the effect of cache polluting attacks
would be to detect them prior to taking action. This may be achieved with
$s(u)$, if a ground truth estimation of its shape exists. So, if $s(u)$ is
known beforehand, an estimate of the average miss rate for the cache size used
can be computed. Then, if the instantaneous miss rate surpasses this estimate
by more than configured threshold, a cache protecting action could be taken. For
instance, the top, or most recently used, part of the cache could be
``protected'' against eviction.
Other measures could include the implementation of a cache hierarchy or the
limiting of user request rate for new destinations. In the former case,
evicted entries should be stored in a larger but higher access time cache
while in the latter some of the network elements should monitor per user traffic
and filter out attack attempts. Since it is not within the scope of this paper we
do not explore other aspects related to the implementation of such tools.

\section{Related Work}\label{sec:rw}
Feldmeier~\cite{feldmeier:rt_cache} and Jain~\cite{jain:dst_locality} were
among the first to evaluate the possibility of performing destination address
caching by exploiting the locality of traffic in network
environments. Feldmeier analyzed traffic on a gateway router and showed that
caching of both flat and prefix addresses can significantly reduce routing
table lookup times. Jain however performed his analysis in an Ethernet
environment. He was forced to concede that despite the locality observed in
the traffic, small cache performance was struggling due to traffic generated
by protocols with deterministic behavior. Both works were fundamental to the
field however their results were empirical and date back to the beginning of
the 1990s, years when the Internet was still in its infancy. 

Recently, Kim et al.~\cite{kim:rcaching} showed the feasibility of route
caching after performing a measurement study within the operational
confinement of an ISP's network. They show by means of an experimental
evaluation that Least Recently Used (LRU) cache eviction policy performs close
to optimal and that working-set size is generally stable with time. We also
observe the stability of the working-set for our data sets but we further
leverage it to build a LRU model instead of just empirically evaluating its
performance. 

Several works have previously looked at cache performance in location/identity split
scenarios considering LISP as a reference implementation. Iannone et
al.~\cite{iannone:lcache} performed an initial trace driven study of the LISP
map-cache performance. Instead of limiting the cache size by using an eviction
policy, their cache implementation evicted stale entries after a configurable
timeout value. Further, Kim et al.~\cite{jkim:lcache} have both extended and
confirmed the previous results with the help of a larger, ISP trace and by
considering LISP security aspects in their evaluation. Ignoring control plane
security concerns, which we did not consider, and despite differences
regarding the cache eviction policy, the results of these last two works seem
to be in agreement with ours. Zhang et al.~\cite{zhang:lcache} performed a
trace based mappings cache performance analysis assuming a LRU eviction
policy. They used two 24-hour traffic traces captured at two egressing links
of the China Education and Research Network backbone network.  They concluded
that a small cache can offer good results. Finally, Jakab et
al.~\cite{jakab:lisp-tree} analyzed the performance of several LISP mapping
systems and, without focusing on a cache analysis, also observed very low miss
rates for a cache model similar to that used in~\cite{iannone:lcache}.

Our work confirms previous LISP cache analysis results however, it also tries
to provide a better understanding of the reasons behind the relatively good
performance of map-caches. In this sense it introduces an analytical model
that could be used to theoretically evaluate or dimension for operational
needs the caching performance. Moreover, to the best of our knowledge, it is
also the first work to perform an analysis and propose an analytical model for
the map-cache performance when under scanning data-plane attacks.

\section{Conclusions}\label{sec:conclusions}
The implementation of a location/identity split at network level has been
recently recommended as a viable solution to the Internet's routing scalability
problems. But, unless deployed with care, the cure may prove worse than the
disease. Hence, a good understanding of the newly introduced components, like
the map-cache, is paramount.

In this paper, we propose a methodology to evaluate map-cache performance.
Our model is built by exploring the link between cache performance and
parameters that approximate the intrinsic locality of packet level user
traffic. To this end, we advance the use of the working-set model as a tool to
capture said properties but also as a performance predictor. Accordingly, we define a
framework wherein to perform the analysis and find that the clustering of the
working-set curves is the only condition needed to ensure the
accuracy of the model.
We empirically validate our result by emulation, using traffic traces
collected at the edges of a campus and an academic network.   

Besides the possibility of using the model for cache dimensioning or detecting
attacks in operational environments, we believe the equation may also be used
as part of more complex models that evaluate the scalability of loc/id
architectures. To stress these points, we show the versatility of our
methodology by characterizing map-cache performance for our datasets and by
building an extension that accounts for cache pollution attacks. Our
observations indicate that increasing cache size quickly diminishes miss rates
in normal conditions but has little to no effect under simple cache pollution
attacks. In the latter case, we advise that more complex management strategies
be devised and set in place. In the future, we plan to fully investigate the
sources of locality in prefix level network traffic to better understand their
impact on cache performance.

\section*{Acknowledgements}\label{sec:ack}
We want to express our gratitude to Damien Saucez, Luigi Iannone and Pere
Barlet for their insightful comments. This work has been partially supported by the Spanish
Ministry of Education under scholarship AP2009-3790, research project
TEC2011-29700-C02, Catalan Government under project 2009SGR-1140 and a Cisco
URP Grant.

\bibliographystyle{abbrv}
\small
\bibliography{bib,rfc}
\normalsize
\appendix

\section{Proof of Proposition~\ref{prop:clustering}}\label{proof:cluster}
\begin{proof}
    If $\forall T$ the working-set size $w(t,T)$ is normally distributed,
    and therefore independent of $t$, it follows that the process generating
    $w(t,T)$ is stationary. This in turn implies the stationarity of the
    process generating the reference string and, as a result, necessity is
    proven. It remains to be proved that if the working-set curves are
    generated by the same stationary process then they will tend to cluster.
    In~\cite{denning:ws_properties} it is shown that for a certain window size
    $T$ the distribution of $w(t,T)$ converges to a normal distribution if the
    locality conditions hold. This proves sufficiency. 
\end{proof}

%
%
%
%
%
%
%

\end{document}